\documentclass[aps,showpacs,nobibnotes,showkeys,superscriptaddress,
amssymb,epsfigs,amsfonts,nofootinbib,amsmath,apsfonts]{revtex4}
\usepackage{graphicx}
\usepackage{dcolumn}
\usepackage{bm}
\begin{document}
\title{\Large\bf Masses of Quarks and Leptons in a Supersymmetric GUT.}
\author{B. B. Deo, L Maharana}
\email{lmaharan@iopb.res.in}
\affiliation{Department of Physics , Utkal University, Bhubaneswar-751004, India.}
\author{P. K. Mishra}
\affiliation{Department of Physics, S.C.S. College, Puri-752001,  India}
\date{\today}
\begin{abstract}
The masses and their variation with energy of the elementary particles obtained by 
solving the one loop renormalization group equation (RGE) in the Minimal 
Supersymmetric Standard Model (MSSM)
is considered. A mass about 115 GeV for all the fermions at GUT scale
is  found as the possible solution of RGE. The mass spectra of all the 
fermions are also obtained within acceptable errors. They are shown graphically to 
desend for the top quark but  ascend  for all other fermions from the laboratory 
to the GUT scale. 
\end{abstract}
\maketitle
Only one of all the parameters of the standard model was first predicted in 1974 by Gaillard
and Lee{\cite{Gaillard74}}. Thereafter came the work of the mass matrix ansatz of Fritzsch 
{\cite{Fritzsch77}}. A complete listing of textures and their relevance to experimental 
findings were made by Ramond, Roberts and Ross (RRR){\cite{Ramond93}}. Regarding the mass 
of the top quark, Pendleton and 
Ross{\cite{Pendleton81}}, Faraggi{\cite{Faraggi92}} and some others
\cite{Das01,Parida99,Demir04,Grzadkowski87} 
have predicted the value to be around 175 GeV, even before the top was 
discovered. Here, however, we choose $M_{top}(M_X)=M_F(X)=M_U \simeq$ 115 GeV as 
the only input. 

   The gauge sector of the standard model is characterised by three coupling constants
$g_3, g_2$ and $g_1$ of $SU_C(3)\otimes SU_L(2) \otimes U_Y(1)$, respectively. 
For the three couplings at mass
$M_X=2.2 \times 10^{16}$ GeV{\cite{Einhorn82}} unite to a unified coupling constant 
$\frac{g_U^2}{4 \pi}=\frac{1}{24.6}$. The RG equations for the couplings in the lowest 
order are given by 
\begin{equation}
16 {\pi}^2 \frac{dg_i(t)}{dt}=c_i~g_i^3(t),~i=1,2,3. \label {eq14}
\end{equation}
The coefficients are $c_1 = 6.6, c_2 =1, c_3=-3$.

The running parameter $t$ is defined as $t=\log_e \frac{\mu}{M_Z}$, so that it varies 
from 0 to $\log_e \frac{M_X}{M_Z} \simeq$ 33. The solution of equation(\ref{eq14}) is 
\begin{equation}
\frac{4 \pi}{g_i^2(t)}=\frac{4 \pi}{g_i^2}-\frac{c_i}{2\pi}t \label {eq15}
\end{equation}
Here, $g_i^2=g_i^2(0)$ are the coupling strengths in the electroweak scale $M_Z$. 
Taking the value of
$M_X=2.2 \times 10^{16}$ GeV and $\frac{4 \pi}{g_U^2}=24.6$, we calculate the value of
$\frac{4 \pi}{g_1^2}=59.24$, $\frac{4 \pi}{g_2^2}=29.85$ and $\frac{4 \pi}{g_3^2}=8.85$.
These are consistent with experimental results. Thus the three gauge couplings are descendants
of one coupling constant $g_{\small U}$.

We write the Yukawa sector SUSY RG equations, which had also been written by 
Babu\cite{Babu87} following Georgi
and Glashow, and Cheng et al\cite{Georgi74}, with the masses in units of 175 GeV, which
is taken to be the mass of the top quark,
\begin{eqnarray}
16 {\pi}^2 \frac{dM_F(t)}{dt}&=&A_F M_F^3(t)+\left[Y_F(t)-G_F(t) \right]M_F(t) \label{eq16}\\
&=&A_F M_F^3(t) + Z_F(t) M_F(t), \label{eq17}
\end{eqnarray}
where $A_F$ is 6 for quarks i.e F=1,..,6 and it is 4 for leptons
i.e. F=7,8...,12. The way, the RGE  written above, is critical and crucial. 
The first term is like the first term of gauge sector and $Z_F=Y_F(t)-G_F(t)$. Here the 
the mass in $M_F$ of equation (\ref{eq17}) missing and this mass in $M_F$  
behaves as an external field like the problems of quantum mechanics 
$Y_F$ is a mixed matrix term,
\begin{equation}
Y_F(t)=\sum_H A_{FH}M^\dag_H(t) M_H(t),~~~H=1,2,3...12.\label{eq17a}
\end{equation}

In MSSM, this matrix $A_{FH}$ is given by the 144 elements given below,
\begin{eqnarray}
A_{FH}=
\left (
\begin{array}{cccccccccccc}
 0 & 3 & 3 & 1 & 0 & 0 & 0 & 0 & 0 & 1 & 1 & 1\\
3 & 0 & 3 & 0 & 1 & 0 & 0 & 0 & 0 & 1 & 1 & 1\\
3 & 3 & 0 & 0 & 0 & 1 & 0 & 0 & 0 & 1 & 1 & 1\\
1 & 0 & 0 & 0 & 3 & 3 & 1 & 1 & 1 & 0 & 0 & 0\\
0 & 1 & 0 & 3 & 0 & 3 & 1 & 1 & 1 & 0 & 0 & 0\\
0 & 0 & 1 & 3 & 3 & 0 & 1 & 1 & 1 & 0 & 0 & 0\\
0 & 0 & 0 & 3 & 3 & 3 & 0 & 1 & 1 & 1 & 0 & 0\\
0 & 0 & 0 & 3 & 3 & 3 & 1 & 0 & 1 & 0 & 1 & 0\\
0 & 0 & 0 & 3 & 3 & 3 & 1 & 1 & 0 & 0 & 0 & 1\\
3 & 3 & 3 & 0 & 0 & 0 & 1 & 0 & 0 & 0 & 1 & 1\\
3 & 3 & 3 & 0 & 0 & 0 & 0 & 1 & 0 & 1 & 0 & 1\\
3 & 3 & 3 & 0 & 0 & 0 & 0 & 0 & 1 & 1 & 1 & 0
\end{array}
\right ).\label{eq14a}
\end{eqnarray}
The gauge factor $G_F(t)$, which are the sum of gauge
couplings, are fixed. We take the values from references {\cite{Gotto68}} 
and {\cite{Dimopoulos}}.
\begin{equation}
G_U(t)=\frac{13}{15}g_1^2(t)+3 g_2^2(t)+\frac{16}{3}g_3^2(t)=\sum_{i=1}^3
K_U^i g_i^2(t). \label {eq18}
\end{equation}
$F=1,2,3$ stand for $U$ and they are degenerate electromagnetic gaugewise. Similarly,
\begin{eqnarray}
G_D(t)&=&\frac{7}{15}g_1^2(t)+3 g_2^2(t)+\frac{16}{3}g_3^2(t),~~F=4,5,6, \label {eq19}\\
G_E(t)&=&\frac{9}{5}g_1^2(t)+3 g_2^2(t),~~F=7,8,9, \label {eq20}
\end{eqnarray}
and
\begin{equation}
G_N(t)=\frac{3}{5}g_1^2(t)+3 g_2^2(t),~~F=10,11,12. \label {eq21}
\end{equation}
Here, $K_N^3=K_E^3=0$, as the leptons do not have the strong colour interaction. We shall
need the integrals
\begin{equation}
- \frac{1}{8 {\pi}^2} \int _0 ^t d \tau ~G_F(\tau)=\sum _{i=1}^3 \frac
{K_i^F}{c_i} \log \left(1- \frac{c_i g_i t}{8 \pi ^2} \right), \label{eq22}
\end{equation}
and
\begin{equation}
- \frac{1}{8 {\pi}^2} \int _0 ^{t_X} d \tau ~G_F(\tau)=\sum _{i=1}^3 \frac
{K_i^F}{c_i} \log \left(1- \frac{c_i g_i t_X}{8 \pi ^2} \right)=
\sum _{i=1}^3 \frac{K_i^F}{c_i} \log \frac{g_i^2}{g_U^2}. \label{eq23}
\end{equation}
Deo and Maharana{\cite{Deo1}}, and Deo, Maharana and Mishra{\cite{Deo2}} observed that equation
(\ref{eq17}), which has to be solved for a given fermion, does not contain the coefficients of
the same mass in the matrix $A$. 
This is the reason we can integrate the RGE. 
This fact has probably been overlooked by previous aauthors,
so their result, with the calculational details become cumbersome and the values obtained
take enormous amount of computation.

Thus the present approach gives a simple method of entangling the mass due to 
finding the solution of 12 differential equations. The terms $Z_F(t)$ can be
exponiented away. We introduce a susidiary mass $m_F(t)$ through

\begin{equation}
M_F(t)=m_F(t) exp \left( \frac{1}{16 \pi ^2} \int _0 ^t Z_F(\tau)~d \tau \right), \label{eq24}
\end{equation}
such that $M_F(M_Z)=m_F(M_Z)=m_F(0)$. They satisfy the equation
\begin{equation}
16 \pi^2 \frac{dm_F(t)}{m_F^3}=A_F~ exp \left( \frac{1}{8 \pi^2} \int _0^t
Z_F(\tau)~d{\tau}\right)~dt. \label{eq25}
\end{equation}
They look similar to the gauge sector RG equation (\ref{eq14}).
This can be solved exactly and  integrating equation(\ref{eq25}) from $M_Z$ to $M_X$, i.e., from
$t=0$ to $t=t_X$, we get
\begin{equation}
\frac{8\pi^2} {m_F^3(M_Z)}=\frac{8\pi^2} {m_F^3(M_X)}+A_F \int_0^{t_X} dt~exp~ \left(
\frac{1}{8 \pi^2} \int_0 ^t Z_F(\tau)~d \tau \right). \label{eq26} 
\end{equation}
Putting back the exponential, we get
\begin{equation}
\frac{M_{top}^2(M_Z)}{M_F^2(M_Z)}=\frac{M_{top}^2(M_Z)}{M_F^2(M_X)}~exp~
\left( \frac{1}{8 \pi^2} \int_0^{t_X}Z_F(\tau)~d{\tau} \right) + \frac{A_F}
{8 \pi^2} \int_0^{t_X}dt~exp \left(\frac{1}{8\pi^2} \int_0^t Z_F(\tau)~d{\tau} \right)
. \label{eq27}
\end{equation}
This is the exact one loop solution in this approximation. $M_F^2(M_Z)$ is the
square of the mass of the fermions at $M_Z$. The descent or ascent running of the 
masses from GUT $M_X$ to electroweak $M_Z$, can be obtained by integrating equation
(\ref{eq25}) from $t=t_X$ to $t$. The result which is not given in reference{\cite{Deo1}} is
\begin{equation}
\frac{8 \pi^2 M_{top}^2}{M_F^2(t)}= \frac{8 \pi^2 M_{top}^2}{M_F^2(M_X)}~exp \left(
\frac{1}{8 \pi^2}\int_t^{t_X} Z_F(\tau)~d \tau \right)+A_F \int_t^{t_X} dt_1~exp 
\left( \frac{1}{8 \pi^2} \int_t^{t_1}Z_F(\tau)~d \tau \right). \label{eq28}
\end{equation}
This is also one loop exact in this approach. By solving equations(\ref{eq27}) and
(\ref{eq28}), we can find $M_F(M_X)$ and $M_F(t)$ respectively.

One retains only those terms containing $M_{top}^2(t)$
occuring in any integral with $Y_F(t)$ in the first approximation. For the top case,
 $Y_{top}(t)$ can be set 
equal to zero in the RG equation for the top is a very good approximation. The top mass
is then given by,
\begin{equation}
1= \frac{M_{top}^2(M_Z)}{M_{top}^2(M_X} C_{top} + D_{top}, \label{eq29}
\end{equation}
where, using integrals(\ref{eq22}) and (\ref{eq23})
\begin{equation}
C_{top}=\prod _{i=1}^3 \left(\frac{g_i^2}{g_U^2} \right)^{\frac{K_i^F}{c_i}}
=0.086, \label{eq30}
\end{equation}
and
\begin{equation}
D_{top}=\frac{6}{8 \pi^2} \int _0 ^{t_X =33}dt ~\prod _i^3 \left(1-
\frac{g_i^2 c_i t}{8 \pi^2} \right)^{\frac{K_i^F}{c_i}}=0.802. \label{eq31}
\end{equation}
Putting these values, the original mass of the top, at an energy of $2.2 \times 10^{16}$ GeV
is, 
\begin{equation}
M_{top}(M_X)=M_{top}(M_Z)\left(1-D_{top} \right)^{-\frac{1}{2}} C_{top}^{\frac{1}{2}}=114~
{\text{ GeV}}. \label{eq32}
\end{equation}
This has been Deo-Maharana's \cite{Deo1} result.

The top mass decreases with energy but all the other quarks and leptons, 
starting from the value
at $M_X$ acquire smaller values and become quite light in the electroweak scale. The descent
or ascent equation(\ref{eq28}) describing the `run' can be put in a form like equation
\begin{equation}
\frac{M_{top}^2(M_Z)}{M_F^2(t)}=\frac{M_{top}^2(M_Z)}{M_F^2(M_X)}C_F(t)+D_F(t), \label{eq33}
\end{equation}
where 
\[C_F(t)=a_F(t) \exp \left(\frac{1}{8 \pi^2}\int _0 ^{t_X}Y_F(\tau)~d \tau \right)\]
and  
\[D_F(t)=\frac{A_F}{8 \pi^2}\int _0 ^{t_X}dt_1~b_F(t_1)\exp \left(\frac{1}{8 \pi^2}
\int _t ^{t_1}Y_F(\tau)~d \tau \right).\]
Here,
\begin{equation}
a_F(t)=\prod_{i=1}^3 \left[\frac{\left(1-\frac{g_i^2c_it_X}{8 \pi^2} \right)}
{\left(1-\frac{g_i^2c_it}{8 \pi^2} \right)} \right]^{\frac{K_i^F}{c_i}}, \label{eq34}
\end{equation}
and
\begin{equation}
b_F(t)=\prod_{i=1}^3 \left[\frac{\left(1-\frac{g_i^2c_it_1}{8 \pi^2} \right)}
{\left(1-\frac{g_i^2c_it}{8 \pi^2} \right)} \right]^{\frac{K_i^F}{c_i}}. \label{eq35}
\end{equation}
This is the general result for all the masses.

For the top, we can take $Y_F \to 0$ and calculate the variation of its mass from
$M_U \simeq$ 115 GeV to the top mass 175 GeV using the equation (\ref{eq36}) below
\begin{equation}
M_{top}(t)=\frac{M_{top}(M_X)}{\left[C_{top}(t)+\frac{M_U^2}{M_{top}^2}D_{top}(t)
 \right]^{1/2}}. \label{eq36}
\end{equation}

The matrix $A_{FH}$ of equation(\ref{eq14a}) gives
\begin{equation}
Y_{top}(t)=3 M_c^2(t) +3 M_u^2(t)+ M_b^2(t)+ M_{\nu_\mu}^2(t)+ M_{\nu_e}^2(t)
+ M_{\nu_\tau}^2(t).
\end{equation}
The equation for the top becomes
\begin{equation}
16\pi^2\frac{dM_{top}(t)}{dt~~~~~}=6M_{top}^3 + \left [ Y_{top}(t)-G_{top}(t)\right ]
M_{top}(t).
\end{equation}
For bottom, however,
\begin{equation}
16\pi^2\frac{dM_{bottom}(t)}{dt~~~~~}=6M_{bottom}^3 + 
\left [ Y_{bottom}(t)-G_{bottom}(t)\right ]M_{bottom}(t).
\end{equation}
$M_{bottom}$ is considerably smaller than $M_{top}$. Furthermore,
\begin{equation}
Y_{bottom}(t)= M_{top}^2(t) +3 M_c^2(t)+ 3M_s^2(t)+ M_{\mu}^2(t)+ M_{e}^2(t)+ M_{\tau}^2(t).
\end{equation}
This type of ordering approaches fails to solve the problem of all other fermions. So 
we find a different method presented below to obtain the values of masses and 
their variation with energy.
Let us construct another function for fermions other than the top $B_F(t)$ and $B_{top}(t)$ 
such that
\begin{equation}
8 \pi^2 \frac{d}{dt}\log B_F^2(t)=Z_F(t). \label{eq37}
\end{equation}
In particular,
\begin{equation}
exp \left( \frac{1}{8 \pi^2} \int_0^{t_X}
Z_F(\tau)~d{\tau}\right)=\frac{B^2_F(M_X)}{B^2_F(M_Z)}~~~~~~\texttt{and}~~~~~~~
exp \left( \frac{1}{8 \pi^2} \int _0^t
Z_F(\tau)~d{\tau}\right)=\frac{B^2_F(t)}{B^2_F(M_Z)}.
\end{equation}
Using equations (\ref{eq27}) and (\ref{eq37}), we find,
\begin{equation}
M_F(M_Z) \simeq M_F(M_X) \frac{B_F(M_Z)}{B_F(M_X)}=
M_F(M_X) \exp \left(\frac{-I_F}{16 \pi^2} \right). \label{eq38}
\end{equation}
The integral $I_F$ is
\begin{eqnarray}
I_F&=&\int _0^{t_X}Z_F(t)~dt=\int _0^{t_X}\left(\sum_GA_{FG}M_G^{\dag}(t)M_G(t) 
-G_F(t) \right)dt, \nonumber \\
&=&\frac{1}{2}\int _0^{t_X}
\left(\sum_GA_{FG}M_G^{\dag}(t)M_G(t)-G_F(t)
+\sum_GA_{FG}M_G^{\dag}(-t)M_G(-t)-G_F(-t) \right)dt, \nonumber \\
&=& \frac{1}{4} \int_{-t_X}^{t_X}\frac{dt}{dM_F}dM_F
\left(\sum_GA_{FG}M_G^{\dag}(t)M_G(t)-G_F(t)
+\sum_GA_{FG}M_G^{\dag}(-t)M_G(-t)-G_F(-t) \right), \nonumber \\
&=&\frac{16 \pi^2}{4} \int_{-M_U}^{M_U} \frac{dM_F(t)}{M_F} \frac{\left(
\sum_GA_{FG}M_G^{\dag}(t)M_G(t)-\frac{1}{2}\left( G_F(t)+G_F(-t)\right)\right)}
{\left(A_F M_F^{\dag}(t) M_F(t)-G_F(t)\right)}. \label{eq39}
\end{eqnarray}
Here we have used equation(\ref{eq17}). $M_U$ is defined below.The poles in the integrand 
as can be easily noticed. Set 
\[M_F(t) =M_U e^{i \theta _F(M_F(t)) n_F}= M_Ue^{i \theta _F(t) n_F},\]
where $n_F$ and $\theta_F$ specify a particular Riemann sheet (This $M_U$ should not
be confused with the mass of the U-quark). $n_F$ is a rotational integer.
$t=M_U e^{i \theta _F(t) n_F}$ maps the complex $t$-plane to the inside of a circle
of radius $M_U$. Using this in equation(\ref{eq39}), we get
\begin{equation}
I_F = i \frac{16 \pi^2}{2} \oint n_F d \theta_F(t) \frac{\sum_G A_{FG}
-\frac{G_F(t)}{M_U^2}}{A_F+\sum_G A_{FG}-\frac{G_F(t)}{M_U^2}}. \label{eq40}
\end{equation}
There are poles in the integrand depending on $M_U$ and $G_F$. The factor $G_F / M_U^2$ 
 in the numerator is inconsequential and need be dropped. For the 
quarks, $A_F +\sum _{G=1}^6=6+7=13$,  for the leptons $4+9=13$ and the integral is
nearly a constant. The integer $n_F$, characterising $F$ from the integral,
using $1= \frac{1}{12} \sum_H$ in equation(\ref{eq40}) is in order and we find
\begin{equation}
{\bar{I}}_F=i n_F \frac{16 \pi^2}{2} \frac{1}{12}\sum_H \sum_G A_{AG}\int_{-t_X}^{t_X}
d \theta_H(t) \frac{1}{A_H+\sum_GA_{HG} -\frac{G_H}{M_U^2}}. \label{eq41}
\end{equation}
In the above, one need to take the  averaged $\bar{I}_F$ value over the twelve fermions. 
Retracing the steps and
using $M_H(t)=M_U e^{i \theta_H(t) n_F}$ in equation(\ref{eq40}), we finally get
\begin{equation}
\bar{I}_F \simeq n_F \frac{1}{12}\sum_H \sum_G A_{HG} \int _0 ^{t_X} dt = \frac{1}{12}n_F
t_X \sum_H \sum_G A_{HG}. \label{eq42}
\end{equation}

As $G$ vary from 1 to 6, whereas $F$ will vary from 2 to 12. 
Only the coefficients $A_{HG}$ are needed to calculate 
$\bar{I}_F$ of equation(\ref{eq42}). The coefficients $A_{FG}$ have nonvanishing values for
$F=2,3,...,12$ and $G=1,2,...,6$.

\begin{equation}
\text{For}~F=2,3,...,6:~~~A_{F_1}+A_{F_2}+.~.~.+A_{F_6}=7,~~\text{and}~~
\text{for}~F=7,8,...,12:~~~A_{F_1}+A_{F_2}+.~.~.+A_{F_6}=9, \label{eq44}
\end{equation}
For the twelve fermions, the sum of the coefficients $A_{FG}$
is $(7 \times 5)+(9 \times 6)=89$. The average of $Z$, i.e.,$\bar{Z}$ is given by
\begin{equation}
Z=\frac{89}{12}~~ \text{and}~~ \bar{I}_F=n_F t_X \frac{89}{12}. \label{eq45}
\end{equation}

The masses of all the fermions, due to quark-lepton equivalence, other than the top is
\begin{equation}
M_F(M_Z) \simeq M_F(M_X) e^{-n_F \frac{t_X}{16 \pi^2}\frac{89}{12}}
\simeq\lambda ^{n_F}M_U, \label{eq46}
\end{equation}
where $\lambda = \exp{\left(-\frac{t_X}{16 \pi^2}\frac{89}{12} \right)}=0.219$. This is
the Wolfenstein parameter and is an excellent result in spite of the approximate
estimates. The values of $n_F$ as in the Table-\ref{tab:table1} for leptons are  
8 for electron(e), 4 for muon($\mu$), 3 for tau 
($\tau$), 16 for electron neutrino($\nu_e$), 9 for muon neutrino($\nu_\mu$),
6 for tau neutrino($\nu_\tau$) and that of quarks are 7 for up(u), 6 for down(d), 4 
for strange(s), 3 for charm(c) and 2 for bottom(b).

\begin{center}
\begin{table}[t]
\caption{\label{tab:table1}Identification of Fermions(Leptons and Quarks)}
\begin{tabular}{|c|c|c|c|}\hline
$n_F$&Mass(GeV) & Fermion(Quarks and Leptons) & Expt.(GeV)\\ \hline
2 &5.5&bottom(b)&\\
3 &1.2& charm(c),tau lepton($\tau$)&1.4,~1.7\\
4 &0.264&strange(s), muon($\mu$)&0.1~-~0.23,~0.01\\
6 &0.012&down(d),tau neutrino($\nu_{\tau}$)&0.055~-~0.115\\
7 &0.0028&up(u)&0.003\\
8 & $6\times 10^{-4}$ & electron(e)&$\times 10^{-4}$\\
9 & $1.33 \times 10^{-4}$ & muon neutrino($\nu_{\mu}$)&1.9$\times 10^{-4}$\\
16& $3\times 10^{-9}$& electron neutrino($\nu_e$) &3$\times 10^{-9}$\\ \hline
\end{tabular}
\end{table}
\end{center}

\begin{figure}[b]
\begin{center}
\includegraphics[width=11.1cm,height=6.5cm]{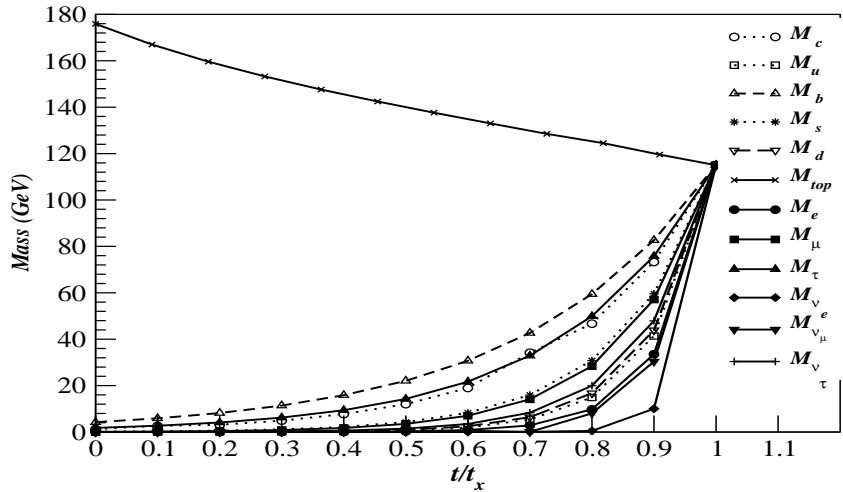}
\caption{\label{fig:Fig5}Variation of Masses(GeV) of all 12 fermions with
t/$t_X$ , t=log($\mu/M_Z$) and $t_X$=33. }
\end{center}
\end{figure}

It is important to note that because of equation(\ref{eq27}), the variation of 
mass with energy comes out automatically also. With 
$Z_F(t)=8 \pi^2 \frac{d}{dt} \log B_F^2(t)$, this reduces to
\begin{equation}
\frac{M_{top}^2}{M_F^2(t)}=\frac{M_{top}^2}{M_U^2}\frac{B_{F}^2(t_X)}{B_F^2(t)}
+\frac{A_F}{8 \pi^2} \int_t ^{t_X} \frac{B_F^2(t_1)}{B_F^2(t)} dt_1. \label{eq49}
\end{equation}
An approximate form of this is to neglect the very small contribution from the 
$A_F$ and we get
\begin{equation}
M_F(t) \simeq M_U \lambda ^{n_F (1-t/33)}. \label{eq50}
\end{equation}
This is shown in the Figure-\ref{fig:Fig5} which shows the variations explicitly.
Much more exact analytic studies are needed to deduce  the exact mass values from the RGE. 
This method does not depend on any previous calculation which do not take both quarks 
and leptons. 
We have succeeded in finding the correct values of the masses and the running masses of the
entire Yukawa masses and we believe that these results will form an important role in 
the physics of particles.

We are very much thankful Dr P K Jena for his help.

\end{document}